\documentclass{elsarticle}
\usepackage{graphicx}
\usepackage{epsfig}
\usepackage{amsmath}
\usepackage{amssymb}
\usepackage{epstopdf}
\usepackage{textcomp} 

\usepackage{lineno}
\linenumbers

\usepackage{multirow}

\usepackage{hyperref}

\def\NIMA#1#2#3{{\rm Nucl.~Instr.~and~Meth.} {\bf{A#1}} (#2) #3}

\begin{document}
\begin{frontmatter}
\title{Data acquisition electronics and reconstruction software 
for directional detection of Dark Matter with MIMAC
}
\author[LPSC]{O.~Bourrion\corref{cor1}}
\ead{olivier.bourrion@lpsc.in2p3.fr}
\author[LPSC]{G.~Bosson}
\author[LPSC]{C. Grignon}
\author[LPSC]{J.~L.~Bouly}
\author[LPSC]{J.~P.~Richer}
\author[LPSC]{O. Guillaudin}
\author[LPSC]{F. Mayet}
\author[LPSC]{D. Santos}

\cortext[cor1]{Corresponding author}
\address[LPSC]{Laboratoire de Physique Subatomique et de Cosmologie,\\ 
Universit\'e Joseph Fourier Grenoble 1,\\
  CNRS/IN2P3, Institut Polytechnique de Grenoble,\\
  53, rue des Martyrs, Grenoble, France}

\begin{abstract}
Directional detection of galactic Dark Matter requires 3D reconstruction of low
energy nuclear recoils tracks. A dedicated acquisition electronics with auto triggering feature and a real 
time track reconstruction software have 
been developed within the framework of the MIMAC project of 
detector. This auto-triggered acquisition electronic uses 
embedded processing to reduce data transfer to its useful part only, 
i.e. decoded coordinates of hit tracks and corresponding energy measurements.  
An acquisition software with on-line monitoring and 3D track 
reconstruction is also presented.

\end{abstract} 
\end{frontmatter}

\section{Introduction}
Directional detection is a promising search strategy of galactic Dark Matter.  
The idea is to take advantage on the rotation of the Solar system around the galactic center to show a direction
dependance of WIMP signals, which should then be clearly discriminated from background ones \cite{spergel,morgan1,morgan2,green1,green2}. 
Recently, a statistical map-based analysis has been developed 
\cite{billard.disco}, showing for the first time the possibility to extract from  data samples of 
forthcoming directional detectors,  both the main direction of the incoming events, thus proving the
galactic origin of the signal,  and the number of WIMP events contained in the map.\\
Several directional detectors are being developed and/or operated :  MIMAC~\cite{MIMAC}, DRIFT \cite{Drift}, NEWAGE~\cite{newage},
 DM-TPC~\cite{mit}. A detailed overview of the status of experimental efforts devoted to directional
dark matter detection may be found in \cite{white}. A common issue amongst these detectors is the fact that directional detection 
requires track reconstruction of recoiling nuclei down to low energies (a few keV). 
This can be achieved with low pressure gaseous detectors \cite{sciolla} and several gases have been suggested : 
$\rm  CF_4,^{3}He+C_4H_{10}$ or  $\rm CS_2$. 
Ideally, recoiling tracks should be 3D reconstructed \cite{green1} with sense recognition \cite{headtail1,headtail2}.\\
The MIMAC (MIcro tpc MAtrix of Chambers) collaboration \cite{MIMAC} is planning to build a multi-target detector,
 composed of a matrix of  gaseous micro-TPC detectors. A pixelized bulk Micromegas \cite{giomataris2} is used in order to perform a 
 3D reconstruction of  few mm tracks. It is segmented in  350\ \textmu m pixels associated to a dedicated ASIC previously described
 \cite{JPRicher}. Indeed, the coordinates in the anode plane 
(x and y coordinates) are reconstructed by collecting primary electrons produced in
the drift region with an electric field ($\leq$1\ kV/cm) and amplified in the avalanche 
region ($20 \rightarrow 80$~kV/cm). The track of the recoil is thus projected on the anode, providing 2D 
information. As stated above, 3D track reconstruction is needed for directional detection of dark 
matter. This is achieved by sampling the anode signal every 25\ ns. Knowing the electron
drift velocity, information on the third coordinate is 
retrieved.\\ 

To demonstrate the relevance of the concept, a specific acquisition electronic 
has been developed in order to equip a prototype detector featuring an anode of $\rm 3.36 \times 3.36\ cm^2$ where 
$2 \times 96$ strips are monitored with these ASICs. This auto-triggered acquisition electronic uses embedded processing to reduce data transfer to its useful part only, 
i.e. decoded coordinates of hit tracks and corresponding energy measurements. 
To be fully exploited, an acquisition software with on-line monitoring and track 
reconstruction has been written.\\

This paper is organised as follows : section 2 presents the hardware design, emphasizing on the front end digital interface (FPGA) implemented on the acquisition board. Then the acquisition software and the 3D reconstruction strategy are presented in section 3. Eventually, experimental results are presented in section 4.

\section{Hardware design and architecture}
\subsection{Overview of the MIMAC acquisition board}
\begin{figure}[th]
\begin{center}
\includegraphics[scale=0.50]{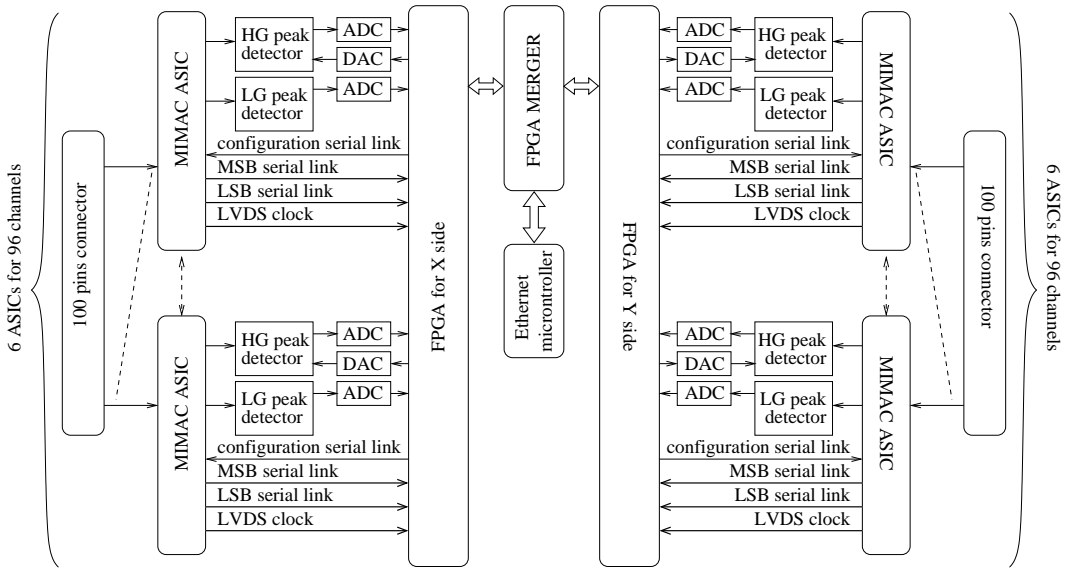}
\caption{Block diagram of the MIMAC acquisition board}
\label{DiagDAQBoard}
\end{center}
\end{figure}

As shown on figure \ref{DiagDAQBoard}, the acquisition board is composed of 12 front end ASICs, 6 for each coordinate (X and Y).
The data processing was split in 3 FPGA, one for each side and one for merging the data. The LVDS serial links are directly connected to the XY FPGA, which de-serializes the ASIC comparators data. Then the XY FPGA processes each ASIC position data for suppressing the zeros, decoding the coordinates and assembling them with the data coming from the neighbouring ASICs. At the end of the processing, the data in the first level FPGA are time sorted and placed in a data buffer (called position buffer).\\
The two analogue outputs of each ASIC go through peak detectors in order to find the maximum amplitude of the shaper signals. Each high gain peak detector is equipped with a comparator whose threshold is tuned by a DAC. The ADC readout is automatically performed by the XY FPGA for each group of 16 strips when some track has been detected and if the amplitude of the signal went above the threshold. As for the position data, the energy data are time sorted and placed in a data buffer.\\
At this stage, for both channels (position and energy), 2 data sources are available: one for the X coordinate and an other for the Y coordinate. The merger FPGA is in charge of concatenating them in a time sorted way and make them available for data readout. An Ethernet ready micro-controller configured as a TCP socket server is used as a communication link with the acquisition computer. It is at the same time used as the slow control interface for:
\begin{itemize}
\item remote configuration of the FPGA (this allows easy upgrades of the firmwares)
\item tuning the energy measurement thresholds
\item tuning the strips discriminators thresholds
\end{itemize}
In the following, we describe the various components of the acquisition board.
\subsection{Front end ASICs}
As a preliminary building block for the MIMAC (MIcro tpc MAtrix of Chambers) framework a front ASIC was developed \cite{JPRicher}. These chips were designed to monitor 16 strips of pixels with low noise charge pre-amplifiers and to provide in real time their time over threshold. 

The comparator outputs are coupled to a data serializer with a compression ratio of 8 in order to reduce consumption and connectivity. Indeed, for the first phase of the MIMAC project, following this prototype phase, the goal is to equip an $\rm 20\ cm \times 20\ cm$ \ anode featuring 1024 strips). De-serializing these data provides a picture of the strips of pixels at a 40~MHz rate. Then, providing the electron drift velocity in the gas mixture the third coordinate of the track is reconstructed by using consecutive samples of the strips of pixels. The thresholds are individually tuned by 5 bit current DACs.\\
For providing a mean to measure the total energy released in the ionisation channel by the incident particle, the sixteen pre-amplifier outputs are summed and provided in two gains via two shapers. More details may be found in \cite{JPRicher}.

\subsection{Energy measurement with peak detectors}
\begin{figure}[th!]
\begin{center}
\includegraphics[scale=0.40, angle=-90]{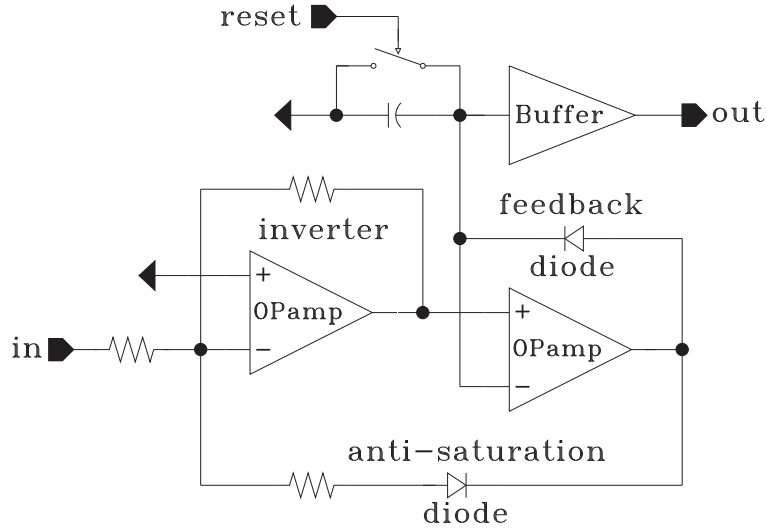}
\caption{Block diagram of the peak detector used for the energy measurement.}
\label{DiagPeak}
\end{center}
\end{figure}

Each ASIC shaper output is connected to a peak detector (fig. \ref{DiagPeak}) which is designed to memorise the maximum amplitude when enabled. Both high and low gain analogical values are digitized by ADCs when the shaper high gain output goes back below the threshold programmed by a DAC (comparator part not shown on fig. \ref{DiagPeak}).
The first stage amplifies and inverts the negative input signal to deliver a positive signal to the second stage. As long as this signal increases, the feedback diode stays in conduction mode and the capacitor voltage follows the signal. Once the maximum amplitude is reached, i.e. when the shaper signal starts to decrease, the feedback diode switches in blocking mode and thus, the peak amplitude is kept in the capacitor. 
At this stage, the peak detector amplifier toggles to non linear mode. An anti-saturation loop is implemented in order to maintain the peak detector amplifier differential input to a minimum value. Indeed one of the bipolar amplifier drawbacks is the fact that they exhibit a dramatic input leakage current increase when input differential voltage increases, which would compromise the analogue memorisation function.
The reset signal is used to disable the peak search and also to discharge the capacitor at the end of the conversion. Bipolar operational amplifiers were chosen for implementation as it was required to work with low voltage supplies and thus having rail to rail input amplifiers was mandatory.

\subsection{Front end digital interface: XY FPGA}
\begin{figure}[th]
\begin{center}
\includegraphics[scale=0.7]{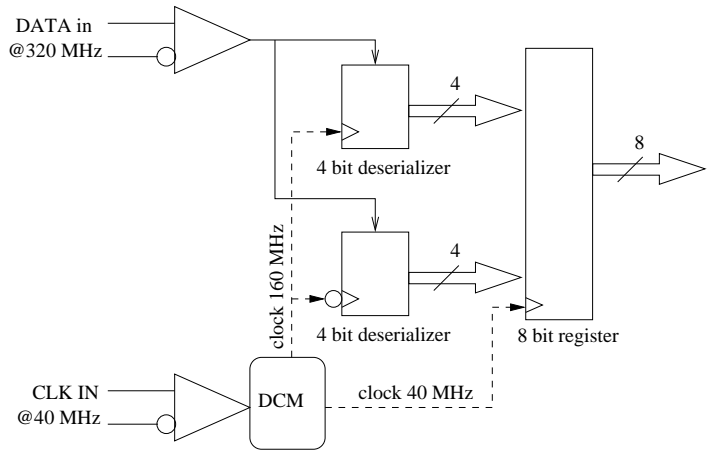}
\caption{Block diagram of the low cost deserializer}
\label{DiagDeserializer}
\end{center}
\end{figure}

\begin{figure}[th]
\begin{center}
\includegraphics[scale=0.6]{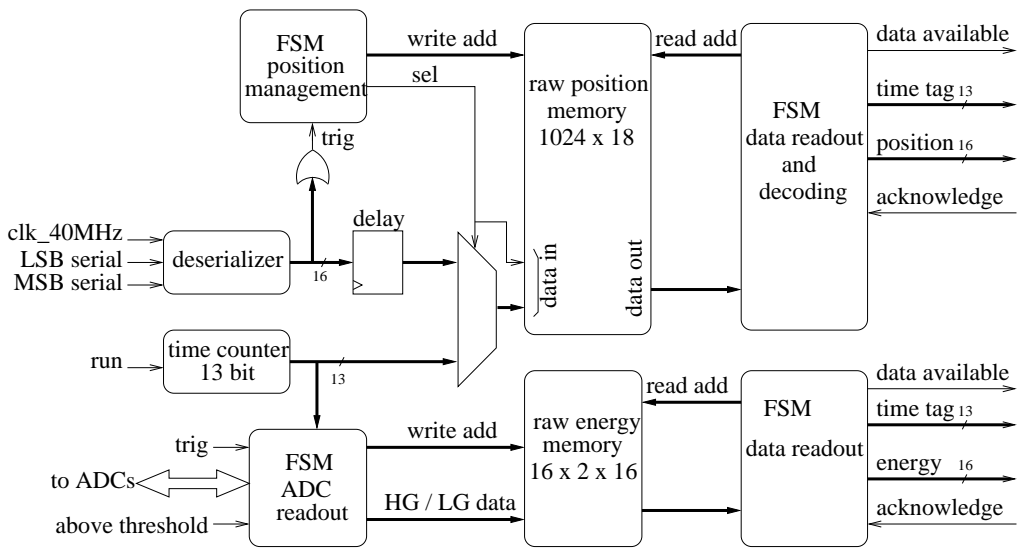}
\caption{Block diagram of the ASIC management logic}
\label{DiagAsicManagement}
\end{center}
\end{figure}

\begin{figure}[h]
\begin{center}
\includegraphics[scale=0.7]{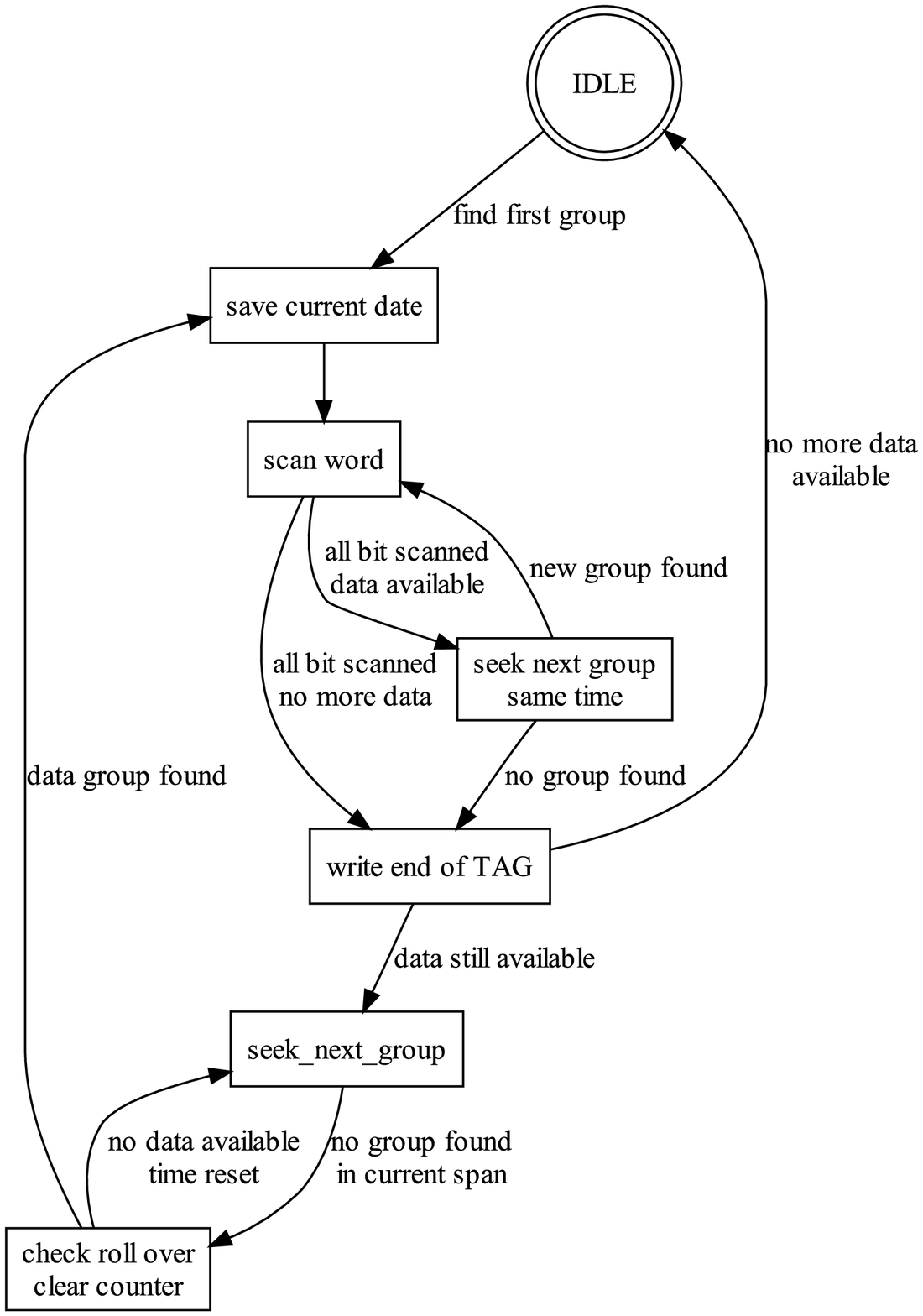}
\caption{Description of the finite state machine (FSM) used for position data sorting}
\label{DiagXYSorting}
\end{center}
\end{figure}

The XY FPGA \footnote{so called because the processing is identical on the X and Y side} is composed of two parts. The first one, identical for each ASIC, deserializes the data and performs the local triggering, ADC readout and local data buffering. The second one is in charge of data time sorting and coordinate decoding.\\
The deserializer is designed to allow operation with low performance FPGA by using two shifters working on opposite edges at half of the incoming data frequency (figure \ref{DiagDeserializer}). The sampling phase is adjusted by setting the DCM (Digital Clock Manager provided in Xilinx FPGAs) to the appropriate value by slow control. \\
The ASIC management is detailed on figure \ref{DiagAsicManagement}. 
The deserializer first provides data for the local triggering (an OR of the 16 channels), when a trigger is found, a copy of the free running time counter is inserted in the \textit{raw position memory}, meanwhile the strip data are delayed by one clock cycle to allow this. After the time count is written in the memory, the strip data are pushed in the memory while the stop condition is not fulfilled. Experimental results show that primary electron distribution can be non continuous and therefore strips can be untriggered for a short amount of time. The stop condition is therefore no trigger on the monitored channel for more than n clock cycles. This value can be adjusted by slow control.\\
The memory depth is a key issue. It has to be dimensioned in order to cope with long tracks like alpha particle, because readout is performed at a much slower speed than the memory writing. For instance, in $\rm ^3He$ at 350~mbar the drift speed is $\rm 16~ \mu m/ns$, so an alpha particle arriving orthogonally to the anode will use $\rm \dfrac{15\ cm}{16~\mu m/ns} \times 40~MHz \simeq 375$ memory cells. Then, each position word in memory can potentially be decoded as 16 valid positions. The memory blocks used in the FPGA are several times larger than this value.\\
The trigger is also used by the energy measurement channel. When detected, the ADC readout system saves the trigger date and waits until a peak is detected within a predefined time limit (shaping time). If none is detected, a null value is inserted in the energy memory in order to keep raw position memory and raw energy memory aligned and facilitate later reconstruction. The raw energy memory is smaller, because for a full particle track only one amplitude will be coded in two gains. Distributed RAM is sufficient for this task.\\
Once all data are stored in local buffers, the next step is to assemble them with the data originating from other ASICs. This operation is done by Finite State Machine (FSM), like the one used for sorting the position data (figure \ref{DiagXYSorting}). The only difference with the one used for sorting the energy data is the position decoding (\textit{scan word} state), each bit at one in position word is decoded as a strip number.
Starting from the \textit{idle} state, the FSM waits for the first ASIC (or group) to present data. When it is the case, the current date is saved and the position decoding is performed. Then the FSM looks if an other group presents position data marked with the same date, if yes the data are decoded and appended to the output buffer in the same time slot, if not the time slot is closed and the search for a new time slot continues. At this stage, 2 possibilities remain:
\begin{itemize}
\item no more data are available and the FSM returns to \textit{IDLE}
\item data are still available and the earliest data has to be found
\end{itemize}
For that, a scan of the dates is made (\textit{seek\_next\_group}) in order to find the first ASIC in time. This scanning is complexified by the fact that the time counter has a short span (13 bit which corresponds to $\sim$1.31~ms), therefore it is performed in 2 stages. First the search is performed from the current date up to the maximum counter value, and if not successful, the current date is set to zero and the searching is performed again until the group is found (\textit{check roll over} state). \\
%
Position data and energy data are stored in separate buffers (see template in table \ref{DataBuffer}). Both tables are organised in order to optimize bit usage and therefore data are encoded on 16 bit. For a specific time-stamp, all useful position or energy data are grouped under a unique TAG. The time slice is closed by the \textit{End of TAG} label.

%
%

\begin{table}[h]
\begin{center}
\begin{tabular}{|c|c|c|c|c|c|c|c|c|c|c|c|c|c|c|c|c|}
\hline
bit  & 15 & 14 & 13 & 12 &11 & 10 & 9 & 8 & 7 & 6 & 5 & 4 & 3 & 2 & 1 & 0 \\
\hline
TAG & 0 & \multicolumn{15}{|c|}{date} \\
\hline
\hline
position DATA n & 1 & 0 &  \multicolumn{7}{|c|}{} & \multicolumn{3}{|c|}{ASIC \#} & \multicolumn{4}{|c|}{strip \#} \\
\hline
\hline
energy DATA n & 1 & 0 & \multicolumn{3}{|c|}{ASIC \#} & \multicolumn{11}{|c|}{ADC value} \\\hline
\hline
\hline
End of TAG & 1 & 1 &  \multicolumn{14}{|c|}{unused}\\
\hline
\end{tabular}
\end{center}
\caption{Organisation of the 16 bit words containing position or energy information.}
\label{DataBuffer} 
\end{table}
\subsection{FPGA merger}
This FPGA is in charge of merging the data from the X and Y side. For this task, it is working similarly to the sorting FSM of the XY FPGA, the only difference being the fact that only 2 data producers are managed instead of 6. Once the data are processed, they are stored in large buffers. They can be accessed directly for readout by the Ethernet micro-controller.
Also, being the direct interface with the micro-controller, it is used as the slow control interface for:
\begin{itemize}
\item Enabling or disabling an ASIC
\item Activating the calibration mode. In this mode a constant pattern is sent by the ASICs and for reducing the data amount, only one out of 65536 triggers is accepted.
\item Configuring the energy channel thresholds
\end{itemize}
\subsection{Ethernet interface}
This interface is based on a 32 bit micro controller, using a flash memory and 16 MB of SDRAM. Its purpose is to provide an easy computer interface and to test the possibility for multi chamber acquisition.\\ 
At power up, a TCP socket server service is started and the acquisition computer (the client) can open a socket and communicate with the hardware via a custom protocol. It is used for slow control and acquisition, they are implemented in 2 threads. When in run mode, the acquisition thread is resumed and start polling the empty flags of the position and energy buffers, once a non empty flag is found, the corresponding fifo level is read. Then the appropriate number of readout is made and the data are transmitted through the socket with a header specifying the data type and count.\\
In order to allow flexibility and easy prototyping, the micro-controller firmware is the only one resident on the electronic board and the 3 FPGA are remotely reconfigured at each start up via the TCP socket.
%
%
%
%
\section{Acquisition software}
The acquisition software was developed in C++ and coded to allow operation on Linux and Windows platforms. In order to keep a good separation between functionalities, it is built with three different parts as described below.
\subsection{Software driver}
The low level driver is in charge of providing low level functions for controlling the hardware and managing the acquisition. It features a platform independent non blocking socket client manager, routine for reading FPGA configuration file and remotely configuring them, methods for interpreting position and energy threshold files and set DAC values.\\
For a proper tuning of the serializers sampling phase, a scanning algorithm is also provided. Basically, it puts the hardware in calibration mode (in order to significantly decrease the trigger rate), validate one ASIC at a time and try every possible value of de-phasing (256 values) and record the results of the reception of a known test pattern. The test pattern is supposed to be received correctly many thousand times in order to consider that the sampling phase is correct for the selected ASIC. When the stability range common to the 12 ASICs has been determined, the  mean value is computed and transferred to the XY FPGA.\\
The data acquisition has to be performed with a blocking call method (returns only when the required number of bytes are read or when a time out occurs). This function reads data frame received from the socket and separates them between energy and position. They are then pushed in 2 STL (Standard Template Library) queues. The readout is performed this way, because the electronic processing time from particle interaction to data available in computer varies with energy and position data path.

\subsection{Acquisition thread and event building}
\begin{figure}[th]
\begin{center}
\includegraphics[scale=0.6]{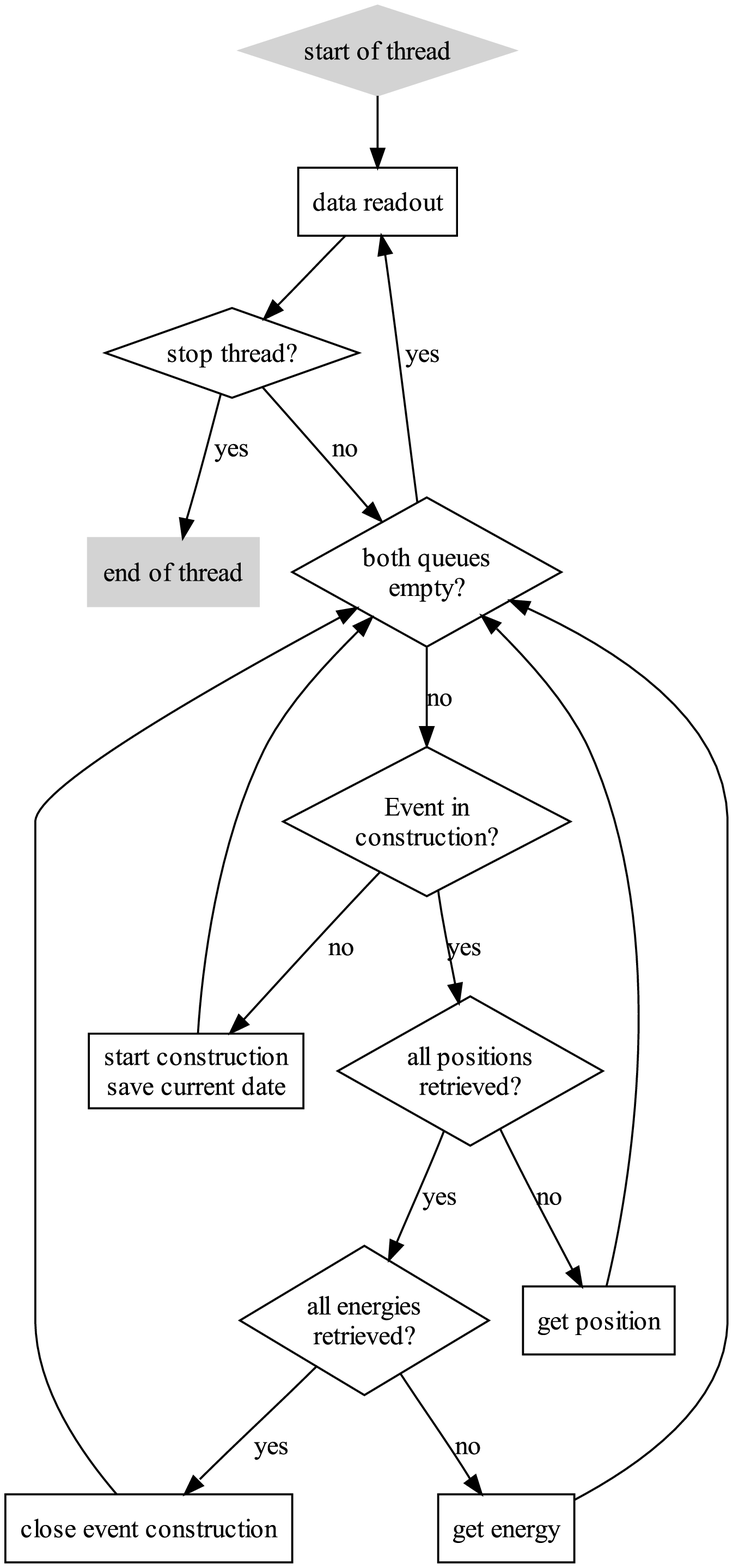}
\caption{Event building algorithm embedded in the acquisition thread}
\label{reconstructAlgo}
\end{center}
\end{figure}

The acquisition is performed in a thread loop, with the software driver and Qt framework \cite{QtFramework}. The algorithm of the event building is depicted on figure \ref{reconstructAlgo}.
When the thread is awaken, it starts looping on the data readout method provided by the software driver. As soon as both position and energy queues are not empty any more the event building processing part starts.\\
In order to properly assemble an event, the position queue is processed first in order to find out when it starts and when it stops. All triggered strips with dates separated by less then the allowed number of empty clock ticks are kept as member of the same event. But as soon as the date of the currently processed triggered strips jumps from more than the allowed number of clock ticks, it is considered to be a member of the next event and the current event position retrieval is finished. At this stage the event start date and duration are known and the corresponding energy data may be found in the energy queue. In other word, the continuous triggering of the anode defines an event.\\

\subsection{Individual strip threshold calibration}
In order to operate correctly a MIMAC prototype equipped with the ASIC described, each strip threshold has to be properly tuned. This task would be cumbersome if not automated, especially when the aim is to monitor 1024 strips per chamber. Therefore an automated calibration method was developed and tested with the prototype.\\
The procedure is to shut down all high voltage applied to the chamber in order to have no real events and to try to find the best DAC setting that will put the thresholds just above noise. This is done by running short acquisitions ($\sim$ 15 s) with each settings until the good one is found.\\
First the calibration algorithm will start with all thresholds set at their maximum values and try to decrease them as long as each individual strip doesn't trigger (all strip thresholds are tuned at the same time). When a strip has reached its lowest threshold, it is marked as such. But since there is some crosstalk on the board, a second phase is necessary. It has been observed that, when other thresholds lower, some noise is generated on the acquisition board and picked again by the strips whose threshold were marked as tuned. So in the second step, the threshold is increased, one digit at a time, until the triggering is stopped. The chamber is declared calibrated when all strip thresholds are marked as having reached their minimum value and no more trigger is counted in the acquisition.

\subsection{Graphical interface and event display}

As stated above, the readout is performed in 2D with strips of pixels, which means that an  event is considered as valid if, for a given time slot, at least one strip of pixels in each dimension (X and Y) is fired.
When such a coincidence happens, the position of the fired pixels can be evaluated as the intersection of the fired X strips and Y strips.
Each time slot gives access to a 2D slice of the track, and each slice following each other will lead to a 3D track.

Once an event is built by the 3D reconstruction algorithm, it has to be displayed with a visualization software both in online and offline modes.
This real-time visualisation software allows to monitor several parameters of the detector during a run, such as the energy deposit, the position of the interaction and the length of the track.
Real-time access to these information offers the possibility to adjust the detector parameters, such as the gas pressure, anode and micro-mesh voltages or the electronic thresholds in order to reach optimized working conditions during an acquisition run.
Furthermore, the event display software enables the testing and debugging of the reconstruction software and the analysis strategy in an offline mode.

To meet these specifications, a visualization tool mainly based on the Qt \cite{QtFramework} and ROOT \cite{root} framework was developed. Qt offers a well documented user interface framework implemented in C++. A third interface layer nammed ``Qt layer'' \cite{fine} was used to build the application software.

\begin{figure}[t!]
\begin{center}
\hspace*{-1cm}
\includegraphics[scale=0.52]{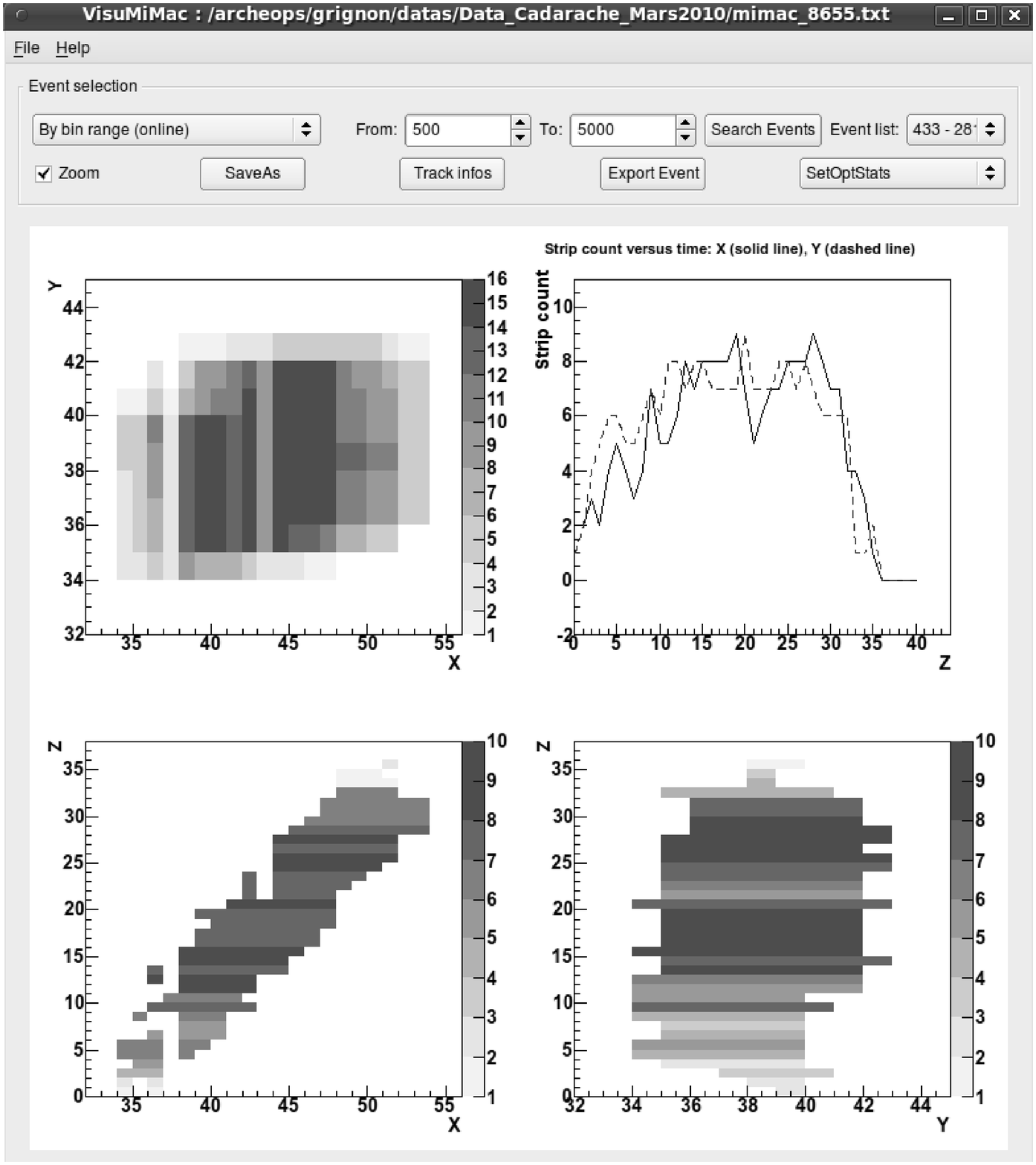}
\caption{Snapshot of the event display tab of the MIMAC real-time visualization software. Top left panel represents a 2D view of the event, as seen by the anode (XY projection). Top right panel shows the evolution of the number of fired strips versus time, for the X side (solid line curve) and the Y side (dashed line curve). Bottom pads represent projections of the track on the XZ plane (left panel) and YZ plane (right panel)}
\label{screenShot}
\end{center}
\end{figure}

The user interface of the online software is segmented in several tabs:
\begin{itemize}
\item one dedicated only to acquisition process: automated calibration, start/stop runs
\item another tab displays the energy spectrum (both in low and high gain) in real time
\item Event display, used also in offline mode, provides projection of the 3D track
\end{itemize}

As shown on figure \ref{screenShot}, the main canvas of the event display panel is segmented in 4 panels:
\begin{itemize}
\item the top left panel represents a 2D view of the event, as seen by the anode (XY projection)
\item the top right panel shows the evolution of the number of fired strips versus time, for the X side (solid line curve) and the Y side (dashed line curve)
\item the bottom pads represent projections of the track on the XZ plane (left panel) and YZ plane (right panel)
\end{itemize}

Event by event display of the projected 3D track (XY, XZ and YZ histograms) is provided by the visualisation software. Whether in online or offline mode, the list of event to display can be adjusted through several criterion like: number of images (equivalent to length along Z axis), energy deposit or even a combination of those.


\section{Experimental results of the MIMAC Data AcQuisition}

As required by the specifications, the MIMAC DAQ is able to reconstruct 3D tracks of nuclear recoils and electrons between a few keV and hundreds of keV.

\begin{figure}[th!]
\begin{center}
\hspace*{-1cm}
\includegraphics[scale=0.35]{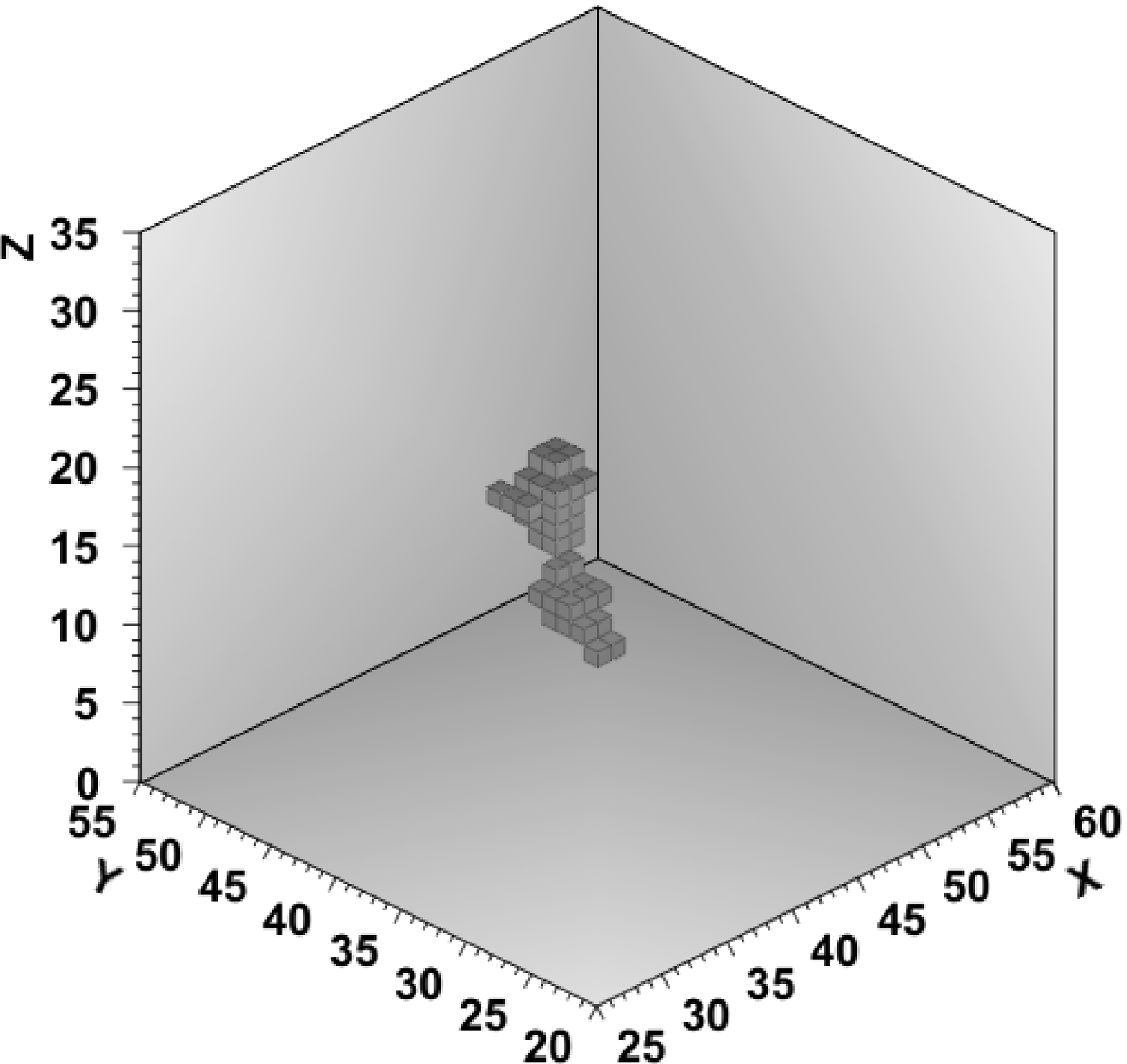}
\includegraphics[scale=0.35]{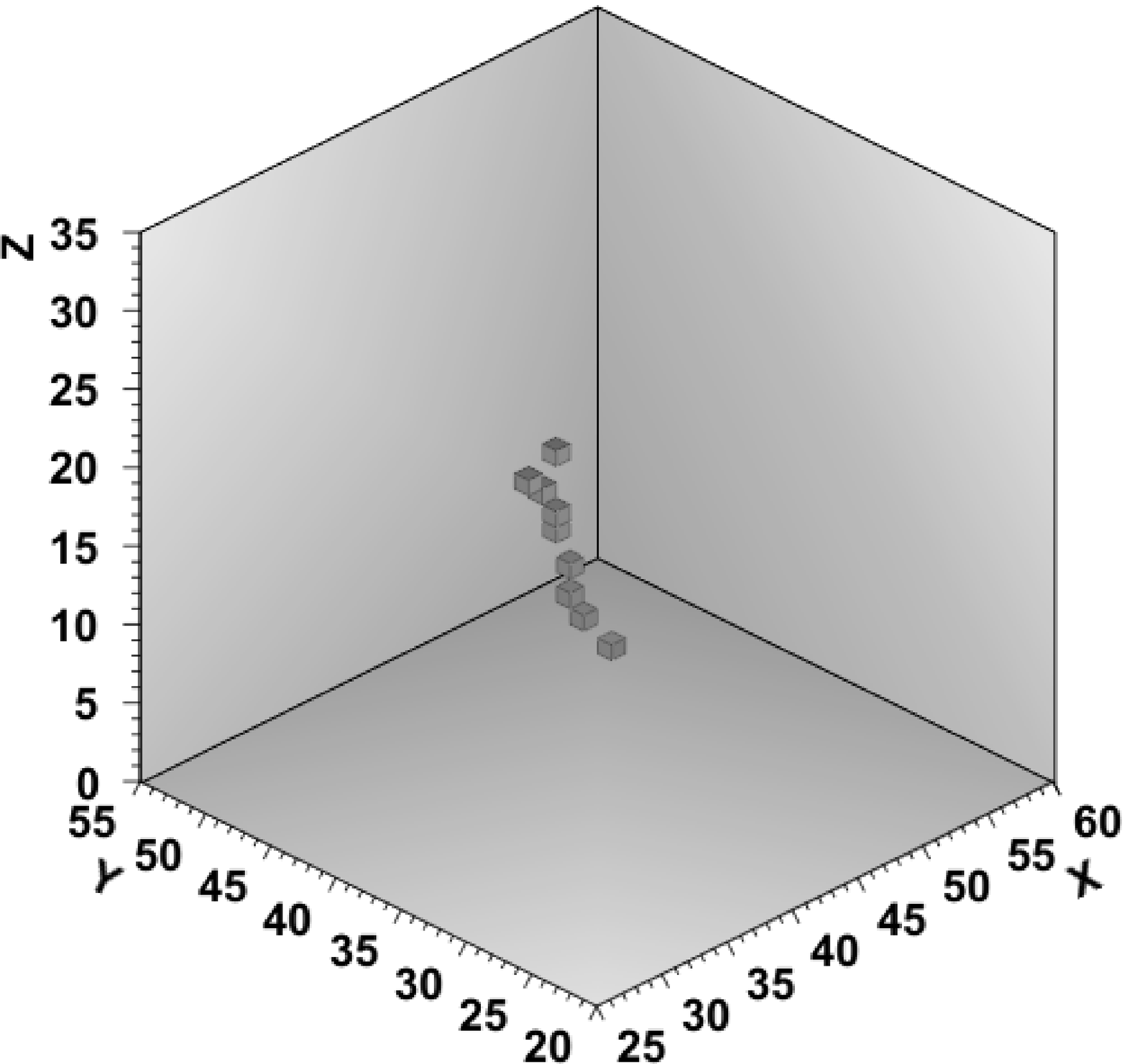}
\caption{5.9 keV electron recoil in a 350 mbar $\rm  ^{4}He + 5\%\  C_4H_{10}$ mixture (barycentre on the right panel)}
\label{electronTrack}
\end{center}
\end{figure}

First experimental results were obtained with a 5.9 keV X-ray source ($\rm ^{55}Fe$) producing 5.9 keV electrons in a 350 mbar $\rm ^{4}He + 5\%\ C_4H_{10}$ mixture used as detection medium.
These electron tracks (figure \ref{electronTrack}) have a specific length over energy ratio and will be typical background events in the final MIMAC dectector designed to detect dark matter in an underground laboratory.

\begin{figure}[th!]
\begin{center}
\hspace*{-1cm}
\includegraphics[scale=0.35]{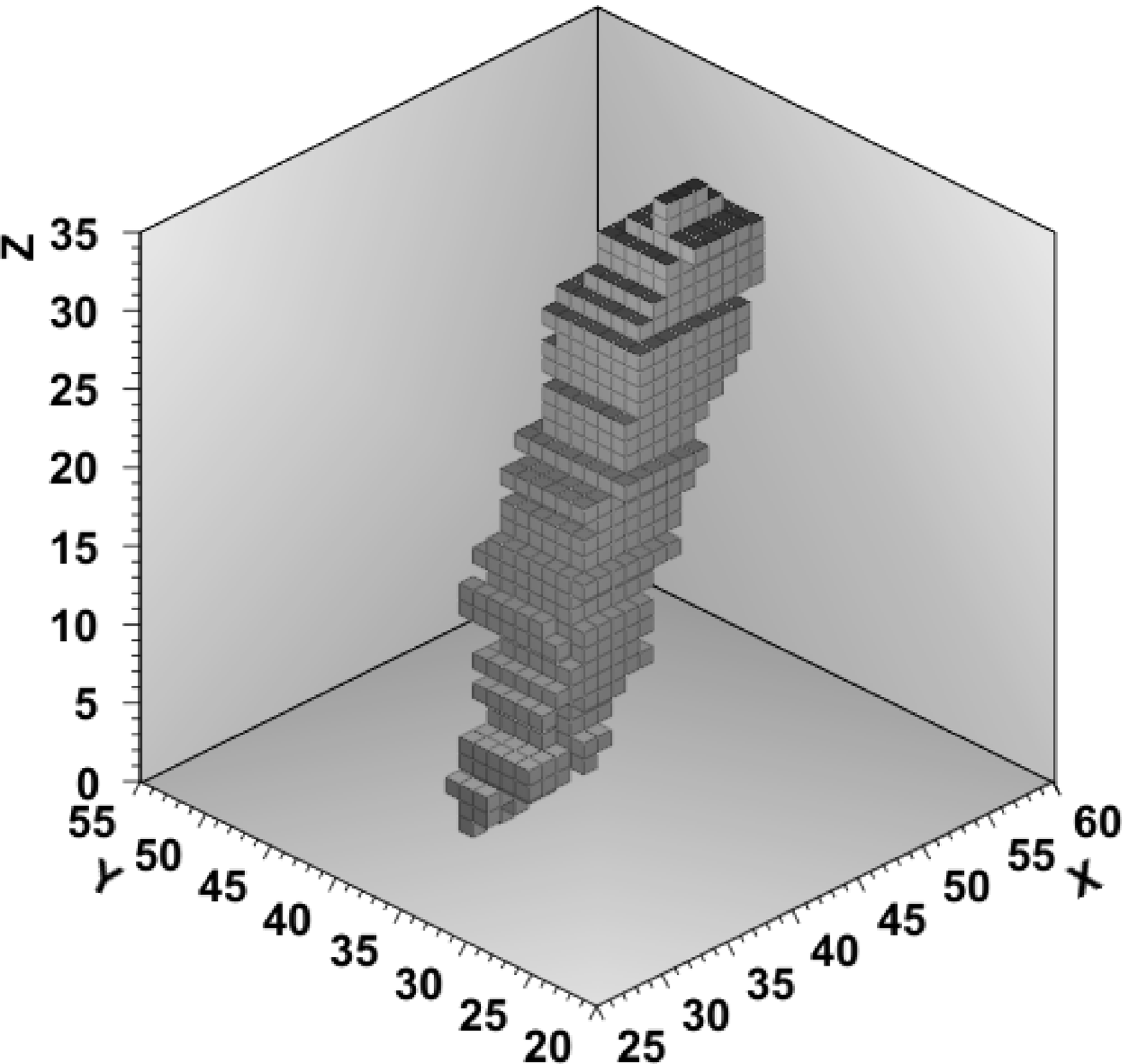}
\includegraphics[scale=0.35]{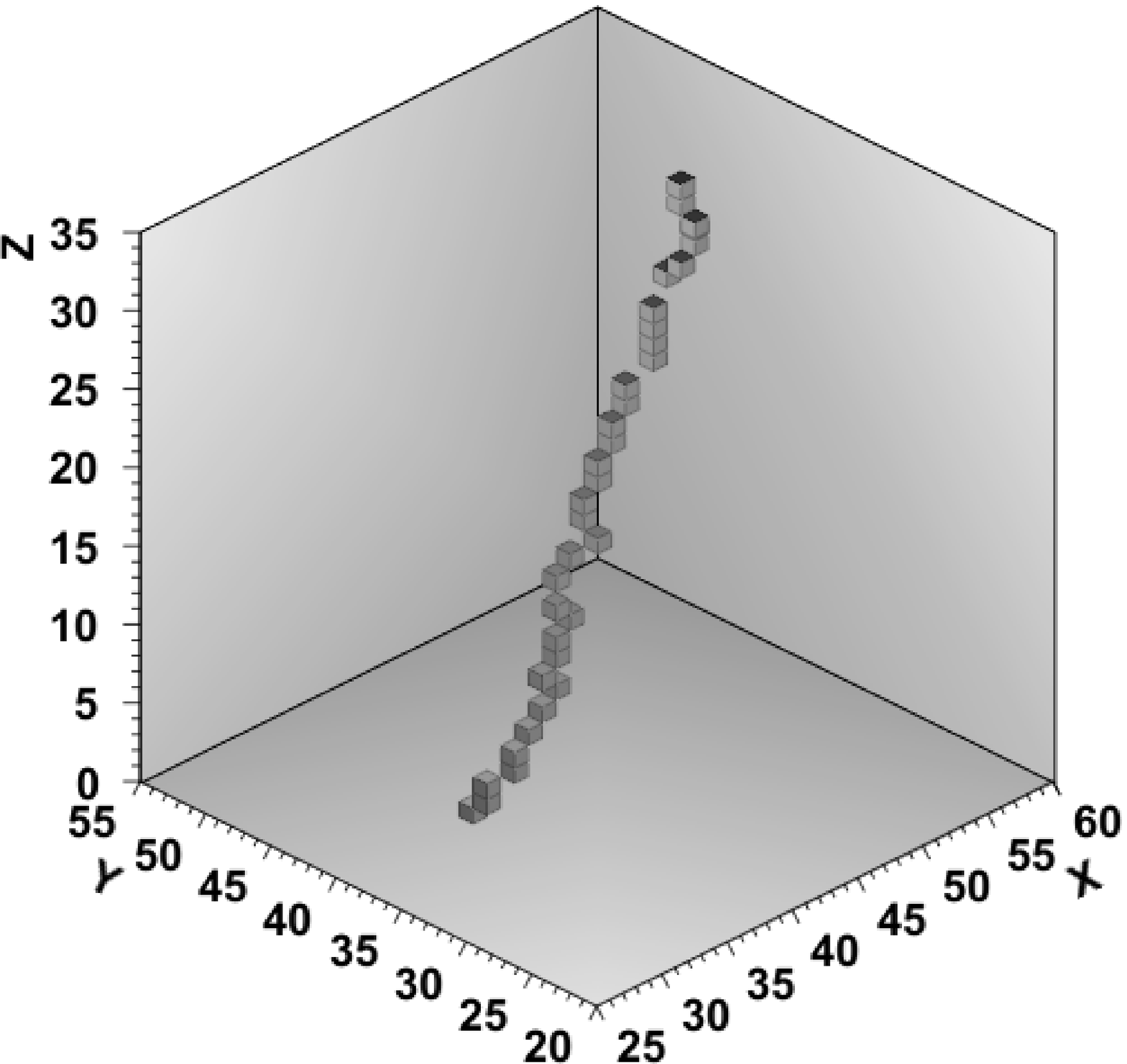}
\caption{100 keV H recoil in a 350 mbar $\rm  ^{4}He + 5\%\  C_4H_{10}$ mixture: left panel represents the raw 3D track while the right panel shows barycentres of each time slice}
\label{protonTrack}
\end{center}
\end{figure}

Ultimately, the goal of MIMAC is to reconstruct recoil tracks of nuclei in 3D.
In order to test the MIMAC prototype associated to the dedicated acquisition electronics with nuclei recoils, the detector was placed in front of a mono-energetic neutron beam at the Amande facility \cite{Amande}. Depending on the target gas used, recoils of hydrogen ions, helium ions or event fluorine ions were reconstructed  \cite{SantosPrep}.

For instance, on figure \ref{protonTrack}, a 100 keV proton recoil obtained in a 350 mbar $\rm  ^{4}He + 5\%\ C_4H_{10}$ mixture is displayed in three dimensions.
Left panel presents the raw 3D track, ie images of the anode in each z slice. Right panel presents the barycentre of each z slice. This shows the possibility to achieve 3D track reconstruction of recoiling nuclei in low pressure gaseous TPC. Figure \ref{electronTrack} presents the same views of 5.9~keV electrons, highlighting the possibility to reach low energies with this technique. This last event is of crucial interest as it features a typical background event for Dark Matter search. A comprehensive study of angular resolutions obtained with this 3D reconstruction method will be presented in a forthcoming paper \cite{GrignonPrep}. This dedicated acquisition electronics associated with 3D reconstruction software offers a major breakthrough towards 3D reconstruction of low energy tracks, thus opening great opportunities for directional Dark Matter search \cite{SantosPrep}.

\section{Conclusion}

\begin{figure}[th]
\begin{center}
\includegraphics[scale=0.75]{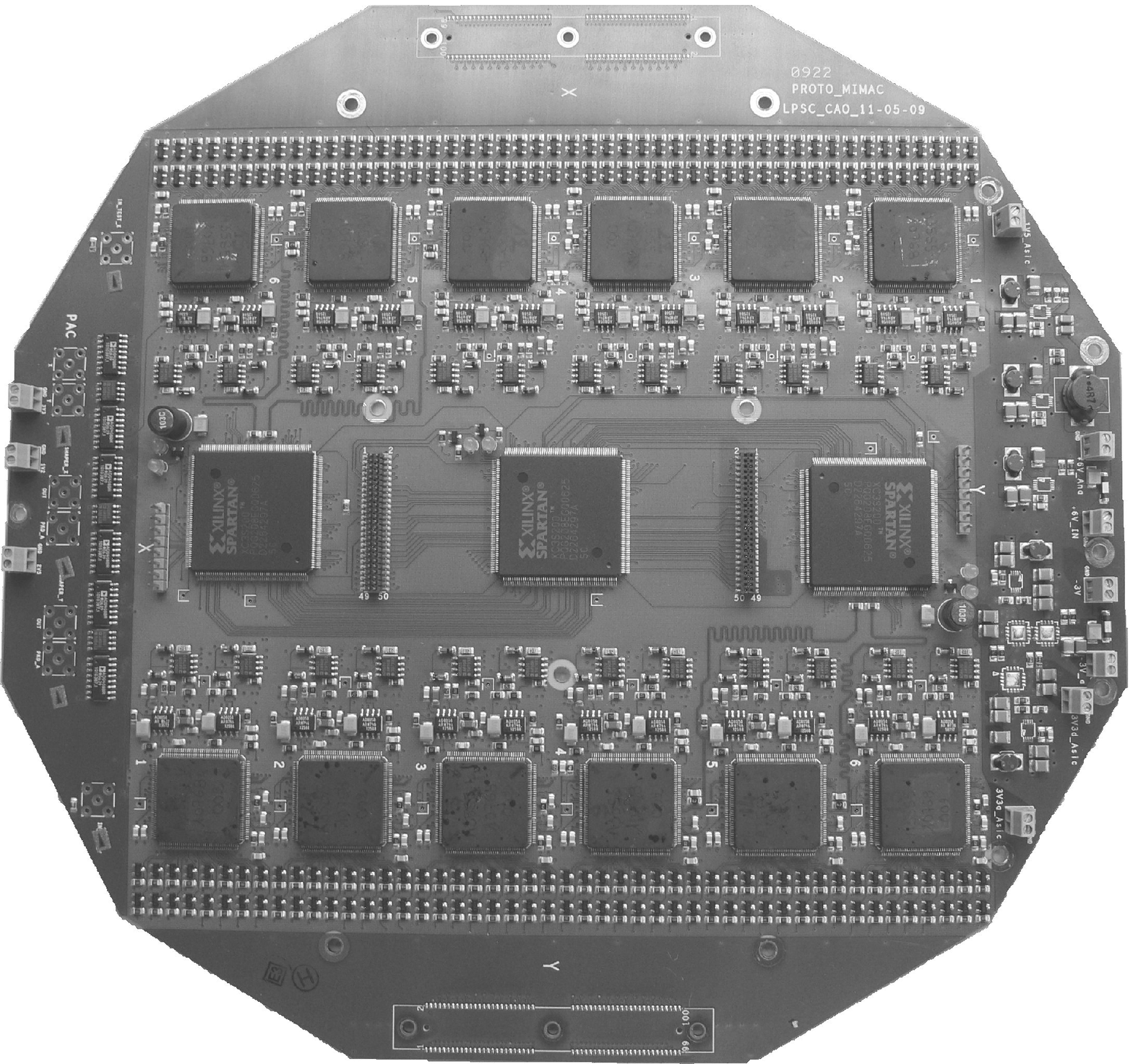}
\caption{Picture of the $\rm 260~mm \times 246~mm$ MIMAC acquisition board. Ethernet interface board and connectors are missing on this view. X side of the dectector is connected on the top connector and Y side on the bottom one. The $2 \times 6$ ASIC are fed by the connectors. In the middle the 3 FPGA perform the processing.}
\label{PictureDAQBoard}
\end{center}
\end{figure}

A dedicated acquisition electronics with auto triggering feature and a real time track reconstruction software have been developed within the framework of the MIMAC project of detector.  It has been shown by experimental results, that the MIMAC ASIC \cite{JPRicher}, the prototype acquisition board (fig. \ref{PictureDAQBoard}) and software coupled together offer great possibilities for 3D track reconstruction and ultimately for directional dark matter detection.\\ 


\section*{Acknowledgement}
C.G. and the MIMAC collaboration acknowledge the ANR-07-BLAN-0255-03 funding.


\end{document}